\documentclass[10pt,preprint2]{aastex}

\usepackage{graphicx}

\begin{document}

\title{Light curve solutions of 12 eccentric \emph{Kepler} binaries and analysis of their out-of-eclipse variability}

\author{Diana Kjurkchieva\altaffilmark{1}, Doroteya Vasileva\altaffilmark{1}
and Dinko Dimitrov\altaffilmark{2}}

\altaffiltext{1}{Department of Physics, Shumen University, 9700 Shumen, Bulgaria}
\altaffiltext{2}{Institute of Astronomy and NAO, Bulgarian Academy of Sciences,
Tsarigradsko shossee 72, 1784 Sofia, Bulgaria}

\begin{abstract}
The eccentricity, periastron angle, orbital inclination, mass ratio,
stellar temperatures and relative stellar radii of 12 eclipsing
eccentric binaries were determined on the basis of \emph{Kepler}
data. The analysis of their out-of-eclipse variability led to the
following results: (i) KIC~10490980 exhibits rotational
(spot-type) variability; (ii) Four new heartbeat stars were found:
KIC~9344623 and KIC~10296163, which have wide tidally induced light humps
and KIC~9119405 and KIC~9673173, which have narrow ''W-shape'' features;
(iii) KIC~4932691 shows oscillations with approximately the 18th
harmonic of the orbital period. We established that the eccentric
\emph{Kepler} binaries fall below the envelope $P(1-e^2)^{3/2}
\approx 5$ days on the period-eccentricity diagram and that there is a
surprising lack of eccentric binaries with periods of 25--35 days.
\end{abstract}


\section{Introduction} \label{sec:intro}

Until half a century ago the eccentric binary systems were objects
mainly of the celestial mechanics. Recently they have become important
targets of the modern astrophysics because their investigation
provides observational tests for the theory of tidal forces and
for the models of stellar structure and evolution (Fuller $\&$ Lai
2011). Eccentric binary systems have been used as probes
for studying the mechanisms for circularization of the
orbits, the synchronization of the stellar rotation with the orbital
motion and complanarity, the impermanent mass transfer occurring close
to the periastron, and apsidal motion (Kopal 1978, Sepinsky et al.
2007, Claret $\&$ Gimenez 2010, Lajoie $\&$ Sills 2011).
Despite of the recent great progress in the study of the
eccentric binary stars, there remain debatable problems: the
dependence of period and eccentricity and its evolutional meaning, the
empirical relations between the stellar parameters of the
eccentric binaries; the existence of highly eccentric binaries with
short periods, etc.

\begin{table*}
\caption{Parameters from the EB catalog: orbital period $P$;
\emph{Kepler} magnitude $m_{K}$; mean temperature $T_m$; widths of
the primary and secondary eclipses $w_{1,2}$ (in phase units);
depths of the eclipses $d_{1,2}$ (in flux units)} \vspace{0.1in}
\begin{scriptsize}
\begin{tabular}{cccccccc}
\hline\hline
Star        &   $P$ [d]     & $m_{K}$   & $T_m$     &  $w_1$    &   $w_2$   &   $d_1$   &   $d_2$    \\
\hline
KIC 4932691 &   18.1120792  &   13.627  &   7109    &   0.0134  &   0.0107  &   0.1114  &   0.0376  \\
KIC 6841577 &   15.5375346  &   14.875  &   5478    &   0.0134  &   0.0174  &   0.2098  &   0.0171  \\
KIC 8610483 &   48.7993681  &   15.05   &   5920    &   0.0072  &   0.0118  &   0.2783  &   0.1224  \\
KIC 8378922 &   43.2633058  &   14.977  &   5438    &   0.0078  &   0.0157  &   0.3381  &   0.2187  \\
KIC 9025914 &   11.3203214  &   16.756  &   5900    &   0.0146  &   0.02    &   0.2344  &   0.0338  \\
KIC 9344623 &   14.7594793  &   13.822  &   6312    &   0.0223  &   0.019   &   0.3367  &   0.3107  \\
KIC 10296163&   9.2967644   &   13.211  &   6229    &   0.0262  &   0.0158  &   0.2057  &   0.0256  \\
KIC 5986209 &   23.7379705  &   15.147  &   5345    &   0.0055  &   0.0126  &   0.3368  &   0.0868  \\
KIC 12217907&   43.2045856  &   14.291  &   5745    &   0.0125  &   0.0066  &   0.2047  &   0.0613  \\
KIC 9119405 &   18.6463233  &   13.729  &   5569    &   0.0163  &   0.0208  &   0.2195  &   0.1881  \\
KIC 10490960&   5.6824111   &   14.3    &   5787    &   0.0327  &   0.0365  &   0.2038  &   0.1973  \\
KIC 9673173 &   21.29474    &   13.731  &   5838    &   0.0136  &   0.0128  &   0.111   &   0.0939  \\
 \hline
\end{tabular}
\end{scriptsize}
\end{table*}

An important step in the field was made by \emph{Kepler}. This mission's hign-accuracy observations have made it possible to detect tidally induced
brightening and oscillations, which were theoretically
predicted by Kumar et al. (1995). The newly discovered
''heartbeat'' stars represent unique laboratories for
investigation of these higher - order effects (Welsh et al.
2011, Burkart et al. 2012, Thompson et al. 2012). Moreover, tidally induced pulsations could explain the driving mechanism of
excitation of certain oscillation modes (Willems $\&$ Aerts 2002).
The study of the Doppler boosting has also became possible
(Bloemen et al. 2011).

According to the analytic model of Kumar et al. (1995) the shape
of the light increase around the periastron is one-peaked for $i <
30^0$ but becomes two-peaked with a central dip (whose depth and
width increase with \emph{i}) for a stepper orbital inclination.
The shapes of the tidally induced features of 16 eccentric
(non-eclipsing) \emph{Kepler} binaries discovered by Thompson et
al. (2012) were varied: with a dip before brightening, with a dip
after brightening, and with a distinct ''W'' or ''M'' shape. We also
found tidally induced brightening for several \emph{eclipsing}
eccentric \emph{Kepler} binaries (Kjurkchieva $\&$ Vasileva 2015a,
Kjurkchieva et al. 2016, Kjurkchieva $\&$ Vasileva 2016), but with
a ''$\Lambda$'' (hump) shape instead of two-peaked profile. More
statistics are necessary for investigation of this newly discovered
phenomenon.

This paper presents the results of our study of a sample of 12
\emph{Kepler} binaries with narrow eclipses (widths in the range
0.006--0.036 in phase units), but still allowing acceptable light
curve solutions. The aims of our investigation were (i) to obtain
their orbits and physical parameters and thus to provide new data
for establishment/improvement of the empirical relations between
the parameters of eccentric binaries and (ii) to analyze their
out-of-eclipse variability and to search for tidally induced
effects.

\begin{figure*}
\begin{center}
\includegraphics[width=13.7cm,scale=1.00]{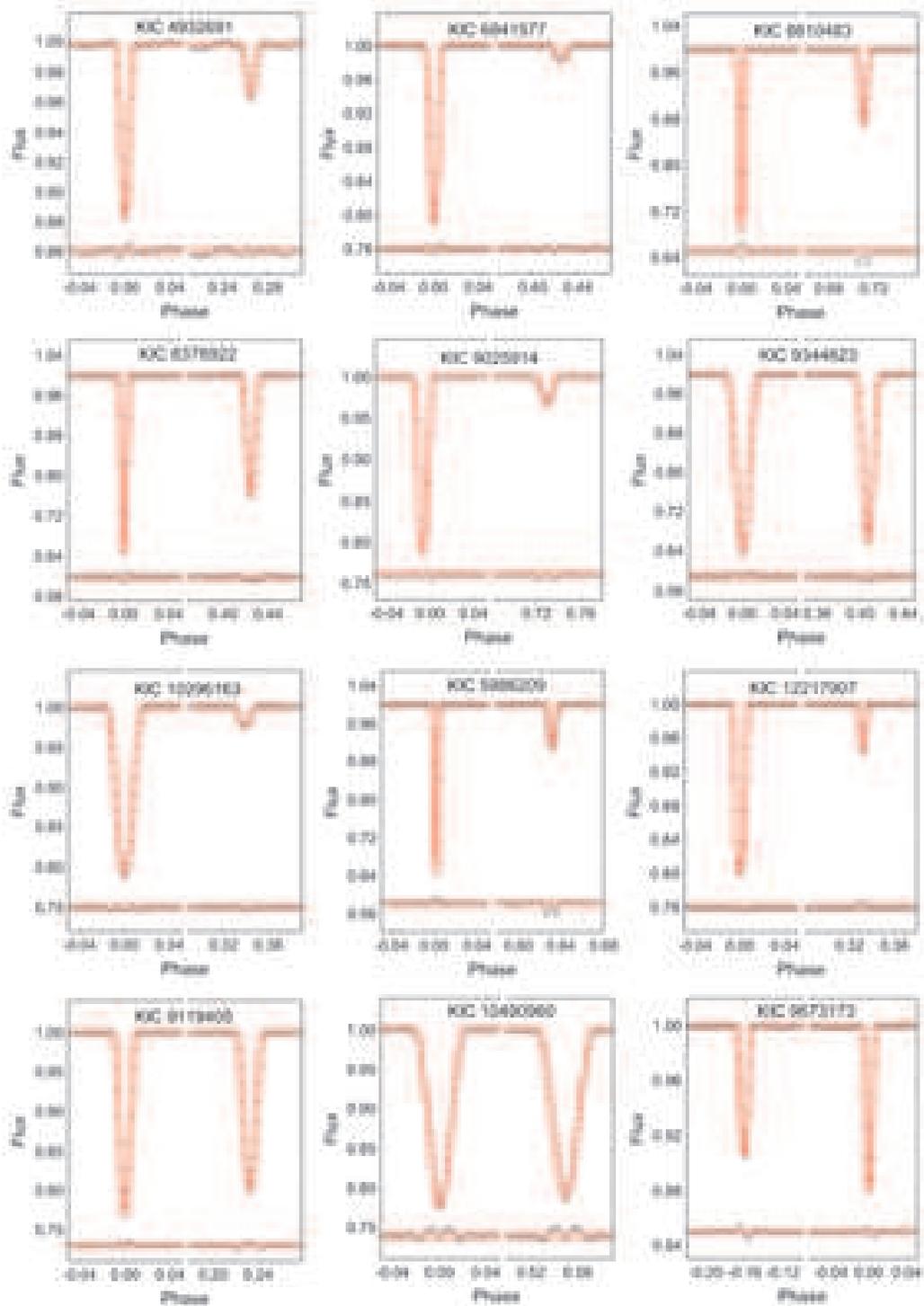}
\caption[]{Top of each panel: the light curve of the target around
the eclipses and its \textsc{PHOEBE} fit; bottom: the
corresponding residuals.} \label{Fig1}
\end{center}
\end{figure*}
\begin{figure}
\begin{center}
\includegraphics[width=8.5cm,scale=1.00]{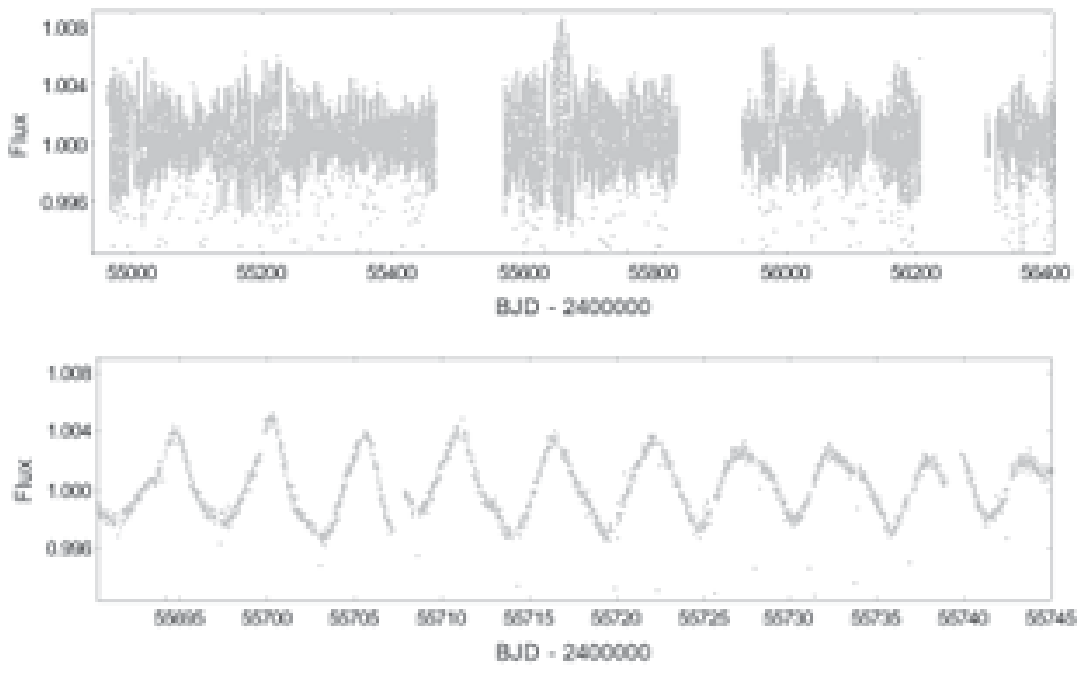}
\caption[]{Out-of-eclipse variability of KIC~10490980 during all
quarters (top) and through 10 consecutive cycles (bottom).}
\label{Fig2}
\end{center}
\end{figure}

\section{Light curve solutions} \label{sec:solutions}

Table 1 presents information for our targets from the EB catalog
(Prsa et al. 2011, Slawson et al. 2011): orbital period $P$;
\emph{Kepler} magnitude $m_{K}$; mean temperature $T_m$; widths of
the primary and secondary eclipses $w_{1,2}$ (in phase units); and
depths of the eclipses $d_{1,2}$ (in flux units). These parameters
and have been obtained by an automated (pipeline) reduction of the
\emph{Kepler} data and/or preliminary ground-based photometric
observations.

Long cadence (LC) data from almost all quarters are available in
the \emph{Kepler} archive for the whole sample. They reveal that
besides eclipses the targets exhibit additional out-of-eclipse
variability (see Section 3). To ignore this effect in the
procedure of the light curve solutions, we modeled all available
photometric data after appropriate phase binning.

We used the package \textsc{PHOEBE} (Prsa $\&$ Zwitter 2005)
for modeling. The procedure is described in detail in
Kjurkchieva and Vasileva (2015a). Briefly, it
consists of several steps: preliminary calculation of the
eccentricity \emph{e} and periastron angle $\omega$; varying of
\emph{e} and $\omega$ to fit the eclipse phases; varying of the
inclination \emph{i}, mass ratio \emph{q}, secondary temperature
$T_{2}$ and potentials $\Omega_{1,2}$ (relative stellar radii
$r_{1,2}$) to reproduce the whole curve; and adjusting the component
temperatures around the mean target temperature $T_m$.

Table 2 shows the values of the orbital and stellar parameters
corresponding to our light curve solutions (the periastron phase
$\varphi_{per}$ and the ratio of relative luminosities
$l_{2}/l_{1}$ were not free but calculated parameters).
The numbers in brackets in Table 2 present the parameter
uncertainties obtained by \textsc{PHOEBE} (rounded to the
greater number). The small errors of the parameters derived by our light
curve solutions are natural consequence of the unique precision of
the \emph{Kepler} data.

The synthetic curves corresponding to our best light curve
solutions are shown in Fig. 1 as continuous lines. The residual
curves show greater scatters during the eclipse phases. Similar
behavior could also be seen also for other \emph{Kepler} binaries
(Hambleton et al. 2013, Lehmann et al. 2013, Maceroni et al.
2014), especially for those with small sum of relative radii.
It has been attributed to the effects of finite integration
time (Kipping 2010), but we found additional reasons for these
discrepancies (see Section 3.1).

The ranges of the parameter values of our targets are as follows
(Table 2):

(a) The temperatures of the stellar components correspond to
spectral type from early F to late K;

(b) The temperature differences of the components reach to around
2000 K;

(c) The mass ratios are in the range 0.32--0.98;

(d) The orbital inclinations are between 86.3$^{0}$ and
89.5$^{0}$. KIC~12217907 and KIC~10296163 undergo total eclipses,
while the other targets show partial eclipses;

(e) The eccentricities of our targets are between 0.1 and 0.56.

\begin{table*}
\caption{Parameters of the best light curve solutions:
eccentricity $e$; periastron angle $\omega$; periastron phase
$\varphi_{per}$ (counted from the primary minimum); inclination
\emph{i}; mass ratio \emph{q}; stellar temperatures $T_{1,2}$;
relative stellar radii $r_{1,2}$; ratio of relative stellar
luminosities $l_{2}/l_{1}$} \vspace{0.1in}
\begin{scriptsize}
\begin{tabular}{cccccccccccc}
\hline\hline
Star        &   \emph{e }  &   $\omega$ [deg] &$\varphi_{per}$ & \emph{i}  &   \emph{q}  & $T_1$ [K]    & $T_2$ [K] & $r_1$ & $r_2$       & $l_{2}/l_{1}$\\
\hline
KIC 4932691 &   0.37886(1)    &    181.7(7)   &   0.14 & 88.06(2)  &   0.393(3)  &   7164(48)   &   5473(28)   &   0.0436(3) &   0.0153(2) &  0.050       \\
KIC 6841577 &   0.12280(3)    &    162.3(4)   &   0.16 & 88.81(3)  &   0.316(2)  &   5560(19)   &   3601(12)   &   0.0391(4) &   0.0176(3) &  0.046      \\
KIC 8610483 &   0.3936(1)     &     32.9(4)   &   0.93 & 88.81(4)  &   0.468(4)  &   6158(14)   &   5653 (3)   &   0.0174(1) &   0.0205(5) &  0.963       \\
KIC 8378922 &   0.2935(1)     &    112.8(2)   &   0.03 & 89.09(1)  &   0.573(5)  &   5548(20)   &   5356(25)   &   0.0229(6) &   0.0234(5) &  0.902       \\
KIC 9025914 &   0.38029(8)    &      5.5(4)   &   0.87 & 88.07(6)  &   0.607(5)  &   6165(19)   &   4327(15)   &   0.0446(2) &   0.0219(1) &  0.052       \\
KIC 9344623 &   0.18902(5)    &    191.5(2)   &   0.31 & 88.99(2)  &   0.658(5)  &   6395(12)   &   6241(12)   &   0.0470(2) &   0.0374(1) &  0.579      \\
KIC 10296163&   0.3674(1)     &    228.2(2)   &   0.29 & 88.53(2)  &   0.401(3)  &   6292(14)   &   4121(18)   &   0.0570(2) &   0.0258(1) &  0.030       \\
KIC 5986209 &   0.3482(2)     &     55.5(2)   &   0.96 & 89.10(3)  &   0.339(8)  &   5503(25)   &   4868(24)   &   0.0188(3) &   0.0144(2) &  0.333       \\
KIC 12217907&   0.38819(4)    &    228.7(1)   &   0.28 & 89.54(3)  &   0.482(6)  &   5850(15)   &   4736(13)   &   0.0253(3) &   0.0111(4) &  0.078       \\
KIC 9119405 &   0.44113(1)    &    168.3(5)   &   0.09 & 88.08(4)  &   0.980(1)  &   5549(22)   &   5573(24)   &   0.0525(4) &   0.0275(3) &  0.280       \\
KIC 10490960&   0.10173(7)    &     41.0(7)   &   0.88 & 86.32(9)  &   0.949(6)  &   5763(34)   &   5838(31)   &   0.0898(7) &   0.0586(6) &  0.449       \\
KIC 9673173 &   0.56510(1)    &      0.8(4)   &   0.92 & 87.22(4)  &   0.752(1)  &   5856(12)   &   5683(14)   &   0.0436(2) &   0.0158(3) &  0.102       \\
 \hline
\end{tabular}
\end{scriptsize}
\end{table*}

\section{Out-of-eclipse variability} \label{sec:out}

All targets exhibit cyclic out-of-eclipse light variability with
the orbital period (excluding KIC~4932691) of different types.

\subsection{Rotational variability}

The out-of-eclipse light curve of KIC~10490980 exhibits long-term
modulation with amplitudes up to 0.01 mag and time scale of
a hundred days (Fig. 2). The tracing of light variability through
several consecutive cycles reveals rotational (spot-type)
variability (Fig. 2). Then the long-term modulation may be
explained by the spot activity cycle of KIC~10490980.

\begin{figure*}
\begin{center}
\includegraphics[width=13.7cm,scale=1.00]{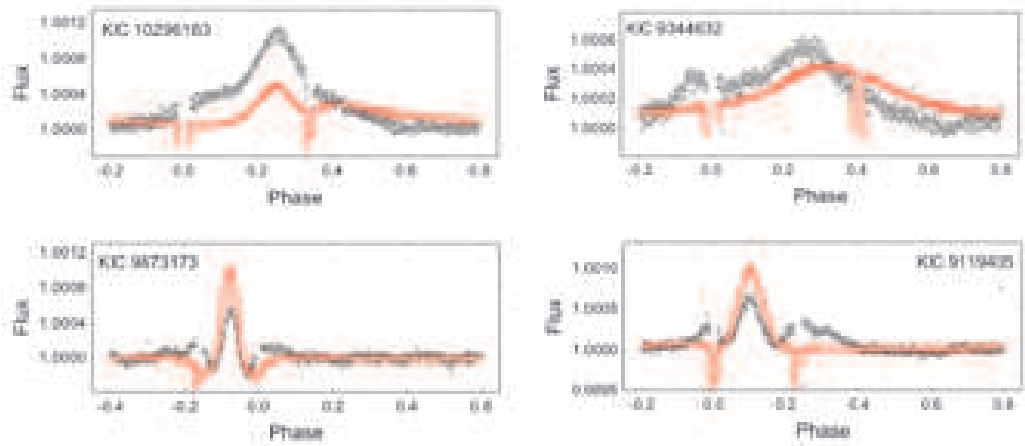}
\caption[]{Folded out-of-eclipse light curves of the targets with
heartbeat effect.}
\label{Fig3}
\end{center}
\end{figure*}

\subsection{New heartbeat stars}

We found tidally induced light brightening with amplitude
0.0004--0.0010 mag around the periastron phase of KIC~10296163,
KIC~9344623, KIC~9673173 and KIC~9119405 (Figure 3).

 The features observed around the periastron of
KIC~9344623 and KIC~10296163 as well as the corresponding
synthetic \textsc{PHOEBE} features are wide and have simple
''$\Lambda$'' (hump) shapes (Fig. 3, top row).

The most interesting features at the periastron phases belong to
KIC~9119405 and KIC~9673173, the targets with the greates
eccentricities from our sample. These features are relatively
narrow and have complex ''W-shape'' profiles (Fig. 3, bottom row).
The synthetic \textsc{PHOEBE} features reproduce the observed ones well (although
not perfectly).

We found that the non-eclipsing star KIC~10873904 from the sample
of Thompson et al. (2012) exhibits similar ''W-shape''
tidally induced brightening with amplitude of 0.001 mag (modeled by
orbital parameters \emph{e} = 0.436, $\omega$ = 347$^0$ and
\emph{i} = 42$^0$).

\subsection{Oscillations of KIC~4932691}

The observed out-of-eclipse light curve of KIC 4932691 (Fig. 4),
the target with the highest temperature, exhibits oscillations
with amplitudes of 0.002--0.004 mag and a period of 1.01626 day (almost 18
times shorter than the orbital cycle). These light variations have a
simple one-waved shape (Fig. 5) and could be considered as
oscillations at the 18th harmonic of the orbital period, i.e. they
could be attributed to the second tidally induced phenomenon
predicted by Kumar et al. (1995).  Most of the heartbeat
targets of Thompson et al. (2012) have oscillations that are
harmonics of the orbital period, but almost all of them
have higher temperatures than our targets. This implies that 
tidally induced oscillations are inherent to early MS stars.

\begin{figure}
\begin{center}
\includegraphics[width=7cm,scale=1.00]{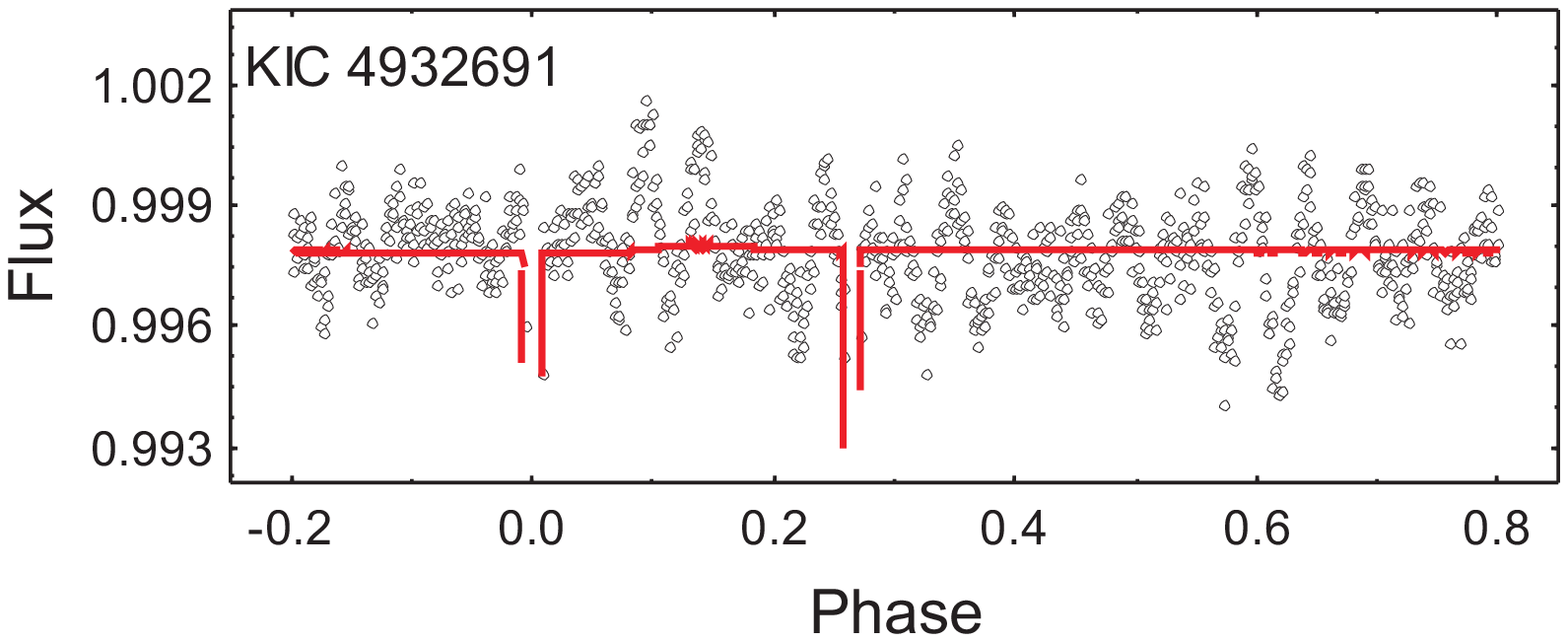}
\caption[]{The oscillations of KIC~4932691.}
 \label{Fig4}
\end{center}
\end{figure}

\begin{figure}
\begin{center}
\includegraphics[width=7cm,scale=1.00]{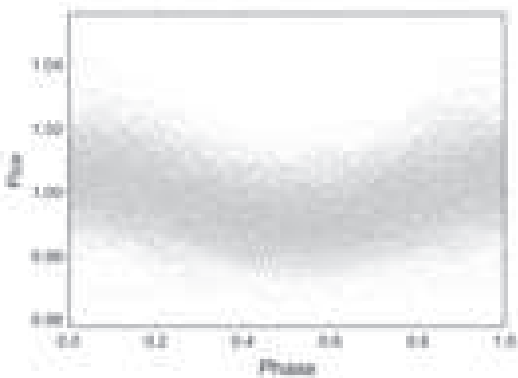}
\caption[]{The out-of-eclipse brightness of KIC~4932691 folded by
period of 1.01626 d.}
\label{Fig5}
\end{center}
\end{figure}

\begin{figure*}
\begin{center}
\includegraphics[width=13.7cm,scale=1.00]{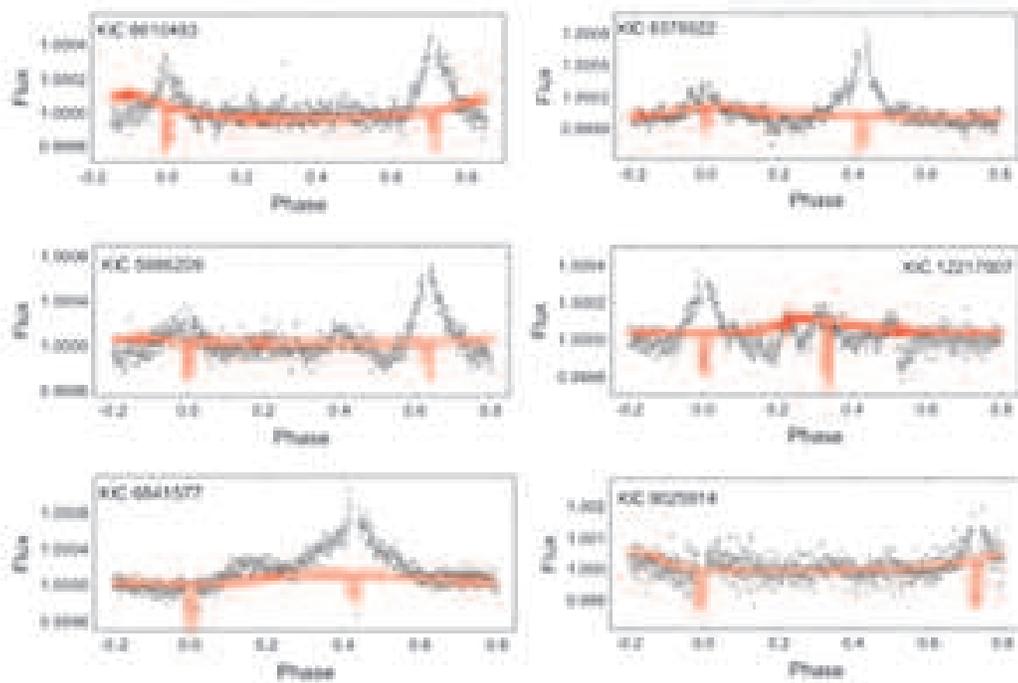}
\caption[]{Folded out-of-eclipse light curves of the targets with
''volcano effect''.}
\label{Fig6}
\end{center}
\end{figure*}

\subsection{Artificial light increasing at the eclipses}

KIC~8610483, KIC~8378922, KIC~5986209, KIC 12217907, KIC~6841577,
and KIC~9025914 reveal strange light increasing around the
eclipses (Figure 6). It is more apparent at the secondary eclipse
(excluding KIC 12217907) and reaches 0.0008 mag. We established
that this ''volcano effect'' was inherent only for the targets
with  the narrowest eclipses ($w_{1,2} \leq$ 0.018).
It turned out, however, that such a feature is not
present in their raw data (Fig. 7). Hence, we suppose
that it is an artificial effect that is a result of the
imperfect guiding of the telescope and the corresponding
de-trending procedure of the \emph{Kepler} data. This supposition
explains why the synthetic \textsc{PHOEBE} light curves do not
reproduce at all the observed light increasing around the
eclipses of these targets (Fig. 6).

\begin{figure}
\begin{center}
\includegraphics[width=6.7cm,scale=1.00]{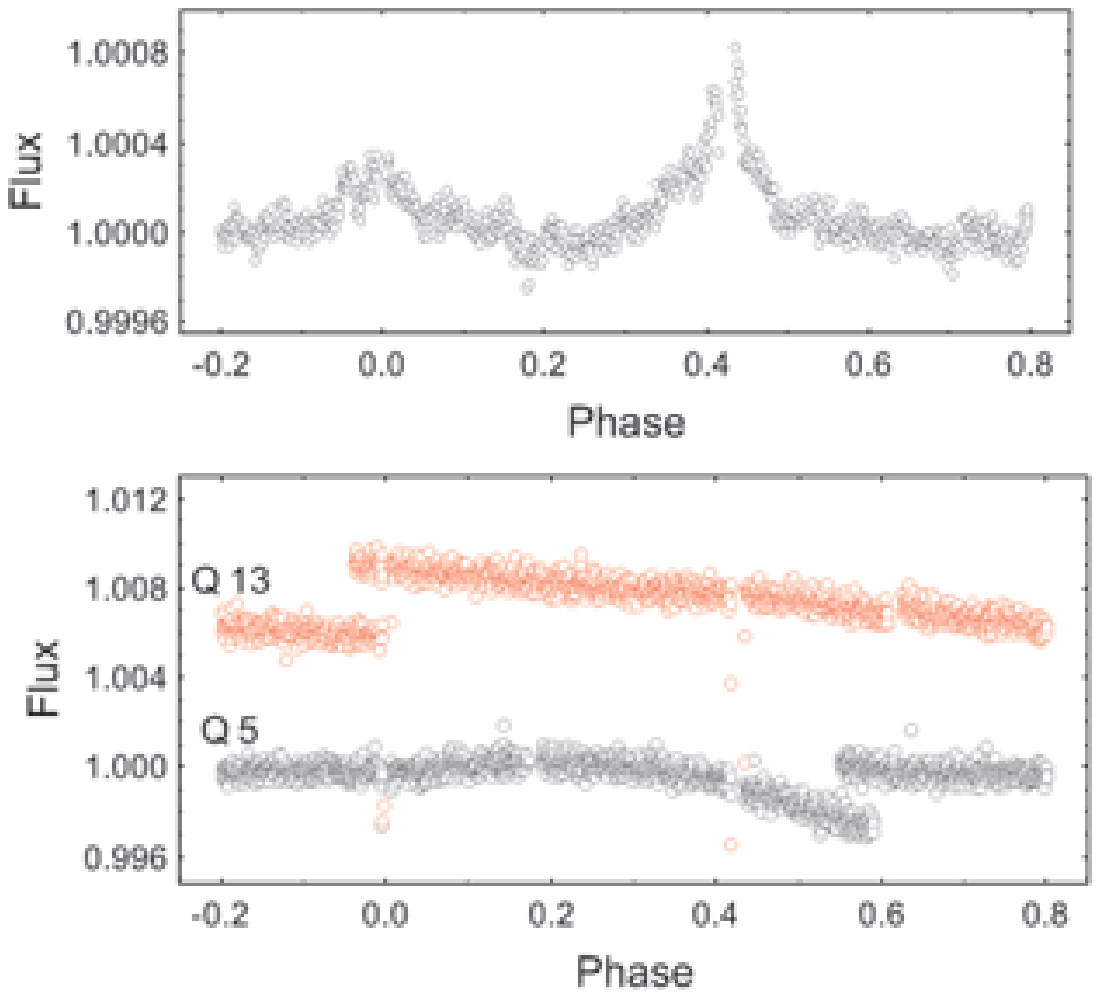}
\caption[]{Out-of-eclipse light variability of KIC~8378922. Top:
the automatically de-trended data with ''volcano effect''; bottom:
phased raw data from quarters Q5 and Q13 (shifted vertically by
0.006).}
\label{Fig7}
\end{center}
\end{figure}

\begin{figure}
\begin{center}
\includegraphics[width=6.7cm,scale=1.00]{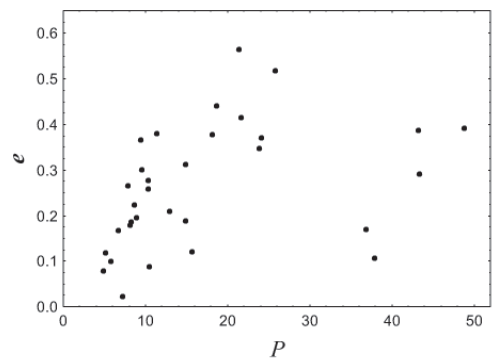}
\caption[]{Period-eccentricity diagram for our 32 \emph{Kepler}
targets.} \label{Fig8}
\end{center}
\end{figure}

\begin{figure}
\begin{center}
\includegraphics[width=7.5cm,scale=1.00]{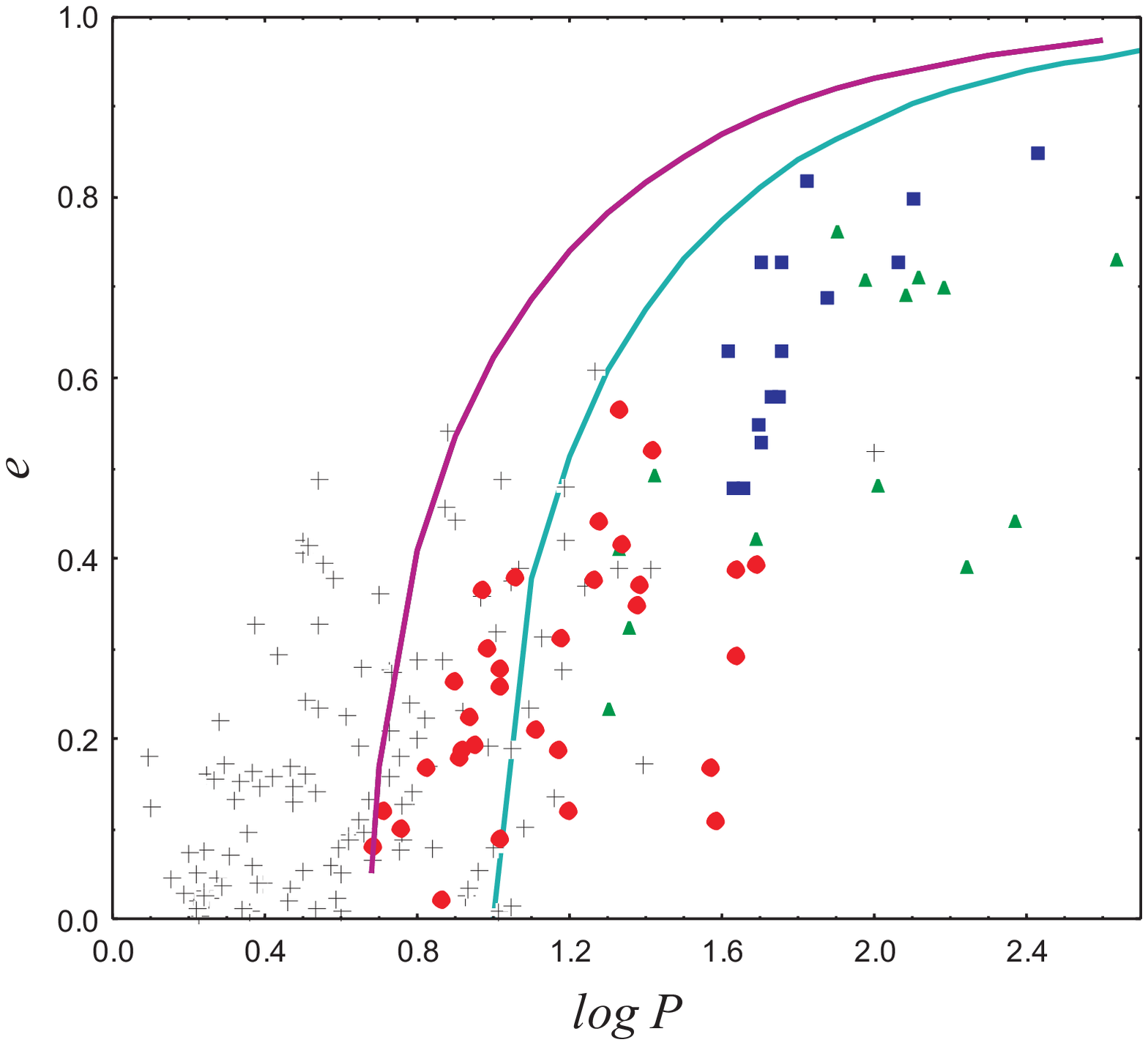} \caption[]{Period-eccentricity diagram of
short-period eccentric binaries: pluses for the sample of Bulut
and Demircan (2007); red circles for our 32 \emph{Kepler}
binaries; green triangles for the \emph{Kepler} targets of Beck et
al. (2014); blue squares for the \emph{Kepler} binaries of Dong et
al. (2013); and continuous green and violet lines describing tracks of
constant angular momentum $P(1-e^2)^{3/2}$ corresponding to 10
days and 5 days.}
\label{Fig9}
\end{center}
\end{figure}

\section{Location on the diagram period-eccentricity}

One of the most interesting and debatable relations of the
eccentric binaries is period-eccentricity. Different attempts have
been made in the past to explain the observed trend that
the binaries with longer periods have greater
eccentricities, including studies of tidal action (Russell 1910); secular decrease in
the stellar mass (Jeans 1924) and the effect of encounters (Walters
1932a,b). Zahn (1977) and Lecar et al. (1976) concluded that the
tidal effect is especially effective for the circularization of
orbits of late stars with convective envelopes.

Horrocks (1936) has derived the theoretical dependence of
period and eccentricity $PM^{2}e^{-6}(1-e^2)^{3/2}$ = const, which has
been confirmed by parameters of binaries with periods $P \geq$ 5
days. Mayor $\&$ Mermilliod (1984) studied red dwarfs in open
clusters and found that the orbits of those with periods shorter
than 5.7 days are circular, while binaries with longer periods have
eccentric orbits. Mathieu et al. (1990) obtained similar results
for 8 evolved MS binaries belonging to old open cluster M67, but
determined a cut off period of around 11 days.

Duquennoy and Mayor (1991) revealed that binaries
with $P \leq$ 10 days have eccentricities close to zero, while
those with longer periods exhibit complex ($e, P$) diagram.
According to Raghavan et al. (2010) the period-eccentricity
relation of solar-type stars shows the expected circularization
for periods below 12 days (caused by tidal forces over the age of
the Galaxy), followed by a roughly flat distribution for 10 $ \leq
P \leq $ 1000 days. The sample of Raghavan et al. (2010) is
limited by an upper envelope $P(1-e^2)^{3/2}$ = 10 days while, the
fifteen highly eccentric \emph{Kepler} binaries studied by Dong et
al. (2013) appear between the tracks of constant angular
momentum corresponding to 10 days and 30 days.

The places on the period-eccentricity diagram of all 32
\emph{Kepler} binaries we have so far studied (Kjurkchieva $\&$ Vasileva 2015a, (Kjurkchieva $\&$ Vasileva
2015b, Kjurkchieva et al. 2016, Kjurkchieva $\&$ Vasileva 2016)
reveal the following trends (Figure 8): (i) There is a fast
increase of the eccentricity (up to 0.57) with the period for
targets with periods of 4 days $ < P < 25$ days; (ii) there is a
second increase of eccentricity beginning from \emph{e} =
0.1 for periods $P \geq$ 35 days; and (iii) there are no
targets with periods within 25--35 days, which may mean the existence of a
short stage of evolution of binaries connected with a
fast change of their orbits.

We built a common ($P, e$) diagram for all \emph{Kepler} eccentric
binaries (Fig. 9). The targets of Dong et al. (2013) and Beck et
al. (2014) fall below the envelope $P(1-e^2)^{3/2} = 10$ days,
while 14 targets (with periods below 10 days and eccentricity $e >
$ 0.1) from our common sample are outside of the region limited by the
foregoing envelope. In fact, all our binaries fall below the
envelope $P(1-e^2)^{3/2} \approx 5$ days (Fig. 9). Most of
the members of the catalog of Bulut and Demircan (2007)
containing above a hundred eccentric binaries appeared on the left
of the last envelope (Fig. 9).

Analysis of the common ($e, P$) distribution of the known
eccentric binaries led us to several results.

(a) There are many short-period binaries (with periods of several
days) with considerable eccentricities. Hence, a possible
conclusion is that there is no sharp cut-off of
the ($e, P$) distribution.

(b) The distribution of the eccentric binaries with periods below
260 days reveals a roughly linear trend of the increase of
eccentricity with the period (Fig. 9), which is expected from the
theory of the binary evolution.

(c) There are no targets with periods of 25--35 days among all
(around 180 in number) eccentric binaries shown in Fig. 9.

\section{Conclusions}

The main results of our study are as follows:

(1) We determined the orbital and stellar parameters of 12
eclipsing binaries with eccentric orbits from the \emph{Kepler}
archive. Their eccentricities are between 0.1 and 0.56, and most of
them show partial eclipses (KIC~12217907 and
KIC~10296163 have total eclipses). The temperatures of the stellar
components correspond to a spectral type from early F to late K.

(2) We found out-of-eclipse light variability of different types:
(i) KIC~10490980 reveals rotational (spot-type) variability. (ii)
Four targets show tidally induced light features around the
periastron phase - i.e. they are heartbeat stars. KIC~9344623 and
KIC~10296163 have wide humps, while KIC~9119405 and KIC~9673173
(the targets with the greatest eccentricities) have narrow
''W-shape'' profiles. (iii) KIC~4932691 reveals oscillations with
approximately the 18th harmonic of the orbital period.

(3) The eccentric \emph{Kepler} binaries fall below the envelope
$P(1-e^2)^{3/2} \approx 5$ days on the period-eccentricity
diagram. We established a lack of eccentric binaries with periods of
25--35 days.

Our investigation may be considered as expansion of the study of
the heartbeat stars of Thompson et al. (2012) to the eclipsing
binaries. It requires further, more in-depth analysis and interpretation
in the framework of the evolutional scenarios of binary systems.

\section*{Acknowledgements}

The research was supported partly by funds of project RD
08-81/03.02.2016 of Shumen University. It used the SIMBAD database
and NASA Astrophysics Data System Abstract Service. We worked with
the live version of the \emph{Kepler} EB catalog
(http://keplerebs.villanova.edu/). The authors are grateful to the
anonymous referee for the valuable recommendations and notes.

\end{document}